\documentclass{elsart}
\usepackage{graphicx,amsmath,amssymb,bm}
\usepackage{amsfonts,amssymb,amscd,amsmath}
\usepackage{graphicx}

\DeclareMathOperator{\Det}{Det}

\newcommand{\slashit}[1]{#1 \kern-.45em\slash}

\newcommand{\slashP}{P \kern-.65em\slash }

\begin{document}
\begin{frontmatter}
\title{Diquark Bose-Einstein Condensation and Nuclear Matter}
\author{A. H. Rezaeian\thanksref{AR}} and
\author{H. J. Pirner\thanksref{P}}
\address{Institute for Theoretical Physics, University of Heidelberg,
Philosophenweg 19, D-69120 Heidelberg, Germany}
\thanks[AR]{E-mail: Rezaeian@tphys.uni-heidelberg.de}
\thanks[P]{E-mail: pir@tphys.uni-heidelberg.de}

\begin{abstract}

\noindent
We study a possible transition between symmetric nuclear matter and
the diquark Bose-Einstein condensate (BEC) matter at zero temperature.
We find that chiral restoration transition is first order and
coincides with deconfinement. We investigate various possible coexistence
patterns which may emerge from the first order deconfinement phase
transition by assuming different values for the critical deconfinement
chemical potential. If deconfinement takes place at higher chemical
potential, there exists a mixed phase of nuclear and chirally restored
diquark BEC matter. This coexistence region extends over a large
density region for a bigger diquark BEC or a weaker diquark-diquark
interaction. For model parameters with heavy diquark in vaccum, phase
transition to diquark matter becomes of second-order.  We also show
that in the case of precocious deconfinement, droplets of nucleons and
droplets of chirally restored Bose-Einstein condensed diquarks coexist
surrounded by non-trivial vacuum. We show that a early deconfinement
and a weak repulsive diquark-diquark interaction soften the equation
of state. We propose a scenario in which nuclear matter saturates due
to the formation of the diquark BEC and deconfinement phenomena. In
this picture, instead of repulsive vector-meson exchange the
compressibility of the equation of state is related to a repulsive
diquark-diquark interaction. In general, we emphasize the importance
of a diquark BEC phase at rather low density before quark BCS-pairing
transition.

\vspace{0.5cm}
\end{abstract}
\end{frontmatter}


\section{Introduction}
Exploring the phases of nuclear and quark matter has recently been
subject of intensive investigation both theoretically and
experimentally. The phase structure of QCD is relevant for a variety
of phenomena, from heavy-ion collisions to neutron stars and the
early universe.  In particular, such studies may reveal the existence of
matter in a new state which may be described by the elementary fields of QCD, i.e. quarks and gluons.

One of the most interesting question in QCD phase diagram is to find
the phase boundary between the hadronic and quark phases. It has been
well do\-cumented that at zero baryonic density, chiral restoration and
deconfinement phase transition occur at the same temperature
\cite{la1}.  Our current understanding of QCD phase structure at
finite density is still rudimentary. Ab initio lattice Monte-Carlo
simulation cannot be directly used due to the complexity of the
sampling weight, the so-called sign problem.  On the other hand, there
is no alternative approach to accommodate all relevant
non-perturbative features of QCD and describe both phases in a unified
way. Various lattice simulation techniques have recently been proposed
in order to circumvent the lattice sign problem (for a recent review
see Ref.~\cite{la3}). But all of them still suffer from systematic
uncertainties which made them reliable at very low
chemical potential. At low temperature and large densities, a number
of model calculations appear to agree on a first order phase
transition between nuclear and quark matter
\cite{nqd,de0}. Therefore, one expects that a first order phase transition at finite
density will end in a second order endpoint and then turns into a rapid
crossover at small chemical potential \cite{la3}. If confirmed, this
indicates that as we approach to lower temperature and finite baryonic
density, the phase transition becomes stronger.

At the nuclear matter saturation point the Fermi momentum $k_{F}$ and
the pion mass $m_{\pi}$ are of comparable scale $k_{F}\approx 2
m_{\pi}\approx 260$ MeV. This implies that at the densities of
interest in nuclear physics $\rho_{0}=0.15
~\text{fm}^{-3}=0.45~m^{3}_{\pi}$, pions and chiral dynamics must be
important. The methods of chiral perturbation theory exploit this
observation and have been successful to describe the nuclear matter
properties \cite{cpt}. By estimating the volume occupied by a single
nucleon from its  mean square charge radius
$r_{\text{rms}}=\sqrt{\langle r^{2} \rangle}$, one may conclude that
baryonic matter is dilute up to about $3\rho_{0}$, since it can satisfy
the diluteness condition $r_{\text{rms}}\rho^{1/3}\ll 1$. At higher
density nucleons start to overlap. Nuclear matter is dilute and at the
same time it is strongly correlated since $a\rho^{1/3}\gg 1$, where $a$
is the scattering length.  On the other hand, since the
Fermi momentum exceeds the QCD scale parameter
$\Lambda_{\text{QCD}}\approx 200 $ MeV, multiple scattering between
nucleons on the Fermi surface start to probe distance of the order
$\approx 1~\text{fm}$ and the nucleonic substructure becomes
visible. The Fermi momentum at saturation point is in fact of the
order of $k_{F}\approx \Lambda_{\text{QCD}}$. The confinement scale
can even become one order of magnitude smaller at moderate baryonic
density
\cite{res}.  Therefore, it is legitimate to ask whether confinement
and nuclear matter saturation mechanisms are interconnected.  The
delta-nucleon mass splitting $\Delta\approx 290$ MeV is also
comparable to the Fermi momentum $k_{F}$ at nuclear matter
saturation. It is known that the main mechanism behind the
delta-nucleon mass splitting is related to their different internal
diquark configurations \cite{n-d}. Namely, the delta is made of only
axial vector diquarks while nucleons can be constructed from both the
scalar and axial vector diquarks. This indicates that perhaps diquark
dynamics at nuclear matter saturation point might be also
relevant. The relevance of diquarks as an efficient way to incorporate
the non-perturbative feature of QCD and to simplify the underlying
dynamics has been well-known for years (cf. \cite{dm0,n-d} and
references therein).  For example, baryons properties in vacuum have
been successfully described in the covariant diquark-quark picture
\cite{n-d}. Recently, it has also been shown that nuclear matter
properties can be described by simulating the nucleons as relativistic
bound states of the diquarks and quarks
\cite{rp,bw}.  At high density diquark concept
is also crucial to describe the color superconductivity phenomenon
\cite{arw,br}. On the other hand, if diquarks as bound states exist,
 they can in
principle undergo Bose-Einstein condensation since they are
bosons. Therefore, it seems to be natural to assume that baryonic matter
at some density undergoes phase transition to diquark BEC matter. One of the
motivations behind this paper is to explore such a transition.

Based on the above scaling argument, it seems that the most important
non-perturbative features of QCD might be involved in  the nuclear matter
saturation density.  In this paper we attempt to address such a
possible connection between the nuclear matter saturation properties
and chiral restoration and the deconfinement effect at zero
temperature. We simulate the quark deconfinement at finite density by
allowing that the active degrees of freedom change from nucleons to diquarks at
high density.  This may lead to a first order deconfinement phase
transition, in accordance with indications from earlier studies
\cite{nqd,de0}. We calculate the nuclear and diquark equations of state (EoS) at zero temperature. 
In particular, we will investigate a possible diquark Bose-Einstein
condensate at about the nuclear matter density. Our result leads to a
non-convential scenario for the stability of the nuclear matter.  We
show that the nuclear matter is energetically favored to undergo the diquark
BEC phase. We study the emergence of various coexistence patterns by
allowing the deconfinement to occur at different critical baryonic
chemical potential. We will show that it is possible that: all three
transitions chiral restoration, the deconfinement and the liquid-gas
transition are coincident at zero-temperature.

This paper is organized as follows: In section 2, we first study the
diquark BEC dynamics. In section 3, we introduce the hadronic part of
our model and calculate the nuclear and diquark EoS.  In
sections 4 and 5 we investigate the transition between diquark BEC
phase and nuclear matter phase at high and low baryonic density, respectively. Some concluding remarks are given in
section 6.

\section{Diquark BEC and spectrum of Goldstone modes}
With any weak attractive interaction fermions exhibit superfluidity
characterized by an energy gap from the BCS pairing mechanism. (For
reviews on color superconductivity, see
Refs.~\cite{bub0,shov}). However, if the attractive interaction is
strong, the fermions can form bound bosonic molecules which condense
into the bosonic zero-mode at some critical temperature. This is
called Bose-Einstein condensation (BEC). These two different
mechanisms are in fact two different physical manifestations with the
same order parameter and are expected to be connected by a
crossover\footnote{This phenomenon has been experimentally examined in
various ultracold gases of fermionic alkali atoms using the Feshbach
resonance techniques
\cite{bec2}.}. In the weak-coupling region at high density the
correlation length between two quarks is bigger than the interquark distance or the pair
size. However, one may expect that in strong-coupling region at rather
low density due to large pair fluctuations, the correlation length
becomes smaller than the pair-size and quark matter undergoes the
diquark BEC phase~\cite{bec1}. Therefore, at the density around the
deconfinement transition, the diquark bound states may be the relevant
degrees of freedom.  Here, we are interested in the physics of
low density (strong-coupling region) where the BEC phenomenon is more
relevant.  Therefore, we assume that the diquarks bound states are
already formed and we concentrate on Bose-Einstein condensation. The
diquark matter has been already studied in Ref.~\cite{wbec} but
without taking into account the diquark BEC effect. We use the
following Lagrangian density for a diquark field $\Phi$,
\begin{equation}
\mathcal{L}_{\text{d}}=(\partial_{0}+i\mu)\Phi^{\dag}(\partial_{0}-i\mu)\Phi-\left(\partial_{i}\Phi^{\dag}\right)\left(\partial_{i}\Phi\right)
-m_{d}^{2}\Phi^{\dag}\Phi-g\left(\Phi^{\dag}\Phi\right)^{2}+\dots,\label{ac0}
\end{equation}
where $\mu$ is the chemical potential with respect to the $U(1)$
charge and $m_{d}$ denotes the diquark mass.  The coupling $g$ gives
the repulsive diquark-diquark interaction. We assume that the
higher-order scattering terms can be ignored.  $\Phi$ is a color
$\bar{3}$ complex scalar field which represents the diquark degree of
freedom,
\begin{equation}
\Phi= \left( \begin{array}{c}
0  \\
0  \\
\phi_{0} \end{array} \right)
+ \frac{1}{\sqrt{2}} \left( \begin{array}{c}
\phi_{11}+i\phi_{12}  \\
 \phi_{21}+i\phi_{22} \\
\phi_{31}+i\phi_{32} \end{array} \right), \label{dw}
\end{equation}
where the complex diquark fields are parameterized in terms of a
possible condensate $\phi_{0}=\langle \phi\rangle$ (due to symmetry,
the direction of condensate is at our disposal) and fluctuating fields
$\phi_{\alpha 1}$ and $\phi_{\alpha 2}$ with $\alpha=1,2,3$. We have
chosen the diquark condensate to be in $\alpha=3$ direction.  The
Lagrangian Eq.~(\ref{ac0}) possesses a $SU(3)\times U(1)$ global
symmetry which spontaneously breaks down to $SU(2)\times U(1)_{Q}$ in
the presence of the condensate. This is exactly the same symmetry pattern which underlies the 2SC phase \cite{bub0,shov}. Having substituted the diquark
field parameterization Eq.~(\ref{dw}) into Lagrangian Eq.~(\ref{ac0}),
one can represent the diquark Lagrangian $\mathcal{L}_{\text{d}}$
Eq.~(\ref{ac0}) as a sum of the condensate Lagrangian
$\mathcal{L}_{C}$, free piece $\mathcal{L}_{F}$ with effective masses
$m_{1}$ and $m_{2}$ and the effective interaction part
$\mathcal{L}_{I}$ among various modes $\phi_{\alpha 1}$ and
$\phi_{\alpha 2}$,
\begin{eqnarray}
\mathcal{L}_{d}&=&\mathcal{L}_{C}+\mathcal{L}_{\text{F}}+\mathcal{L}_{I},\nonumber\\
\mathcal{L}_{C}&=&\left[(\mu^{2}-m_{d}^{2})\phi^{2}_{0}-g\phi^{4}_{0}\right],\nonumber\\
\mathcal{L}_{\text{F}}&=&
\frac{1}{2}\phi_{31}\left(\partial^{\mu}\partial_{\mu}-m^{2}_{1}\right )\phi_{31}
+\frac{1}{2}\phi_{11}\left(\partial^{\mu}\partial_{\mu}-m^{2}_{2}\right)\phi_{11}\nonumber\\
&+&\frac{1}{2}\phi_{21}\left(\partial^{\mu}\partial_{\mu}-m^{2}_{2}\right
)\phi_{21} 
+\sum_{\alpha}\Big(\frac{1}{2}\phi_{\alpha 2}\left(\partial^{\mu}\partial_{\mu}-m^{2}_{2}\right )\phi_{\alpha
2}\nonumber\\ &+&\mu \left[\phi_{\alpha 2}\frac{\partial\phi_{\alpha 1}}{\partial t}-\phi_{\alpha 1}\frac{\partial\phi_{\alpha 2}}{\partial
t}\right]\Big),\nonumber\\
\mathcal{L}_{I}&=&-\sqrt{2}\left[m_{d}^{2}-\mu^{2}+2g \phi^{2}_{0}+g\sum_{\alpha}\left(\phi_{\alpha 1}\phi_{\alpha 1}+\phi_{\alpha 2}\phi_{\alpha 2}\right)\right]\phi_{0}\phi_{31}\nonumber\\
&-&\frac{g}{4}\left(\sum_{\alpha}\left(\phi_{\alpha 1}\phi_{\alpha 1}+\phi_{\alpha 2}\phi_{\alpha 2}\right)\right)^{2}, \label{ac1}\
 \end{eqnarray}
with 
\begin{eqnarray}
m^{2}_{1}&=&-\mu^{2}+m_{d}^{2}+6g\phi^{2}_{0},\nonumber\\
m^{2}_{2}&=&-\mu^{2}+m_{d}^{2}+2g\phi^{2}_{0}.\label{em}\ 
\end{eqnarray}
The inverse of free diquark propagator corresponding to Lagrangian
Eq.~(\ref{ac1}) is a $2\times 2$ matrix in ($\phi_{\alpha 1}$,
$\phi_{\alpha 2}$) space for different color directions, and in momentum
space it reads
\begin{eqnarray}
i\mathcal{D}_{1}^{-1}(\omega,p)&=&i\mathcal{D}_{2}^{-1}(\omega,p)=\left( \begin{array}{cc}
\omega^{2}-p^{2}-m^{2}_{2}& 2i\mu\omega  \\
-2i\mu\omega & \omega^{2}-p^{2}-m^{2}_{2}\end{array} \right),\nonumber\\
i\mathcal{D}_{3}^{-1}(\omega,p)&=&\left( \begin{array}{cc}
\omega^{2}-p^{2}-m^{2}_{1}& 2i\mu\omega  \\
-2i\mu\omega &  \omega^{2}-p^{2}-m^{2}_{2}\end{array} \right).\label{prod}\
\end{eqnarray}
The excitation spectrum is found from the diagonization of the
two-point Green function or, equivalently, by solving the conditions
$\Det(\mathcal{D}^{-1}_{\alpha=1,2,3})=0$.  One finds that six degrees of
freedom of the complex anti-triplet diquark field split into a
degenerate pair of eigenvalues for $\alpha=1,2$ and a singlet pair
$\omega_{3\pm}$ for ($\alpha=3$) (we only consider positive
eigenvalues, the negative ones can be treated similarly),
\begin{eqnarray}
&&\hspace{-1cm}\omega_{1,2\pm}=\pm\mu+\sqrt{\mu^{2}+p^{2}+m^{2}_{2}} \hspace{4.5cm} (\alpha=1,2),\nonumber\\
&&\hspace{-1cm}\omega_{3\pm}=\sqrt{p^{2}+\frac{1}{2}\left(m^{2}_{1}+m^{2}_{2}\right)+2\mu^{2}\pm\frac{1}{2}
\sqrt{\left(m^{2}_{1}+m^{2}_{2}+4\mu^{2}\right)^{2}+4(4\mu^{2}p^{2}-m^{2}_{1}m^{2}_{2})}}\nonumber\\
&&\hspace{9cm}(\alpha=3). \label{ds}\
\end{eqnarray}
It is easy to show that in the absence of a condensate $\phi_{0}=0$
the splitting of the three color directions goes away, as it
should. For $\mu=0$ and no symmetry breaking $m^{2}_{d}>0$ excitation
energies are all identical $\omega=\sqrt{k^{2}+m_{d}^{2}}$ as
expected.

The classical potential has a trivial solution $\phi_{0}=0$ and a
 nontrivial minimum
\begin{equation}
\phi^{2}_{0}=\frac{\mu^{2}-m_{d}^{2}}{2g}. \label{minp}
\end{equation}
In the case that $m_{d}^{2}<0$, symmetry breaking would even occur in
vacuum since $\mu^{2}-m_{d}^{2}$ is then greater than zero. We assume,
however, that $m_{d}^{2}>0$ which means no symmetry breaking at zero
baryon density. Only when $\mu^{2}>m^{2}_{d}$ we have condensation and
the trivial $\phi_{0}=0$ is no longer a minimum of potential
energy. The system develops an instability with respect to formation
of a diquark condensate.  In this ground state, the initial
$SU(3)\times U(1)$ symmetry spontaneously breaks down to $SU(2)\times
U(1)$. Using Eq.~(\ref{minp}), the effective masses in Eq.~(\ref{em})
simplify to $m^{2}_{1}=2(\mu^{2}-m_{d}^{2})$ and $m_{2}=0$, and the
dispersion relations in Eq.~(\ref{ds}) become
\begin{eqnarray}
\omega_{1,2\pm}&=&\pm\mu+\sqrt{\mu^{2}+p^{2}},  \nonumber\\
\omega_{3\pm}&=&\sqrt{p^{2}+3\mu^{2}-m_{d}^{2}\pm
\sqrt{\left(3\mu^{2}-m_{d}^{2}\right)^{2}+4\mu^{2}p^{2}}}. \label{ds1}\
\end{eqnarray}
The small momentum expansion of the dispersion relations in far infrared region is given by
\begin{eqnarray}
\omega_{1,2 +}&\backsimeq& 2\mu+\frac{p^{2}}{2\mu},\\
\omega_{1,2-}&\backsimeq& \frac{p^{2}}{2\mu}\label{ss1},\\
\omega_{3+}&\backsimeq& \sqrt{\frac{2\left(3\mu^{2}-m_{d}^{2}\right)^{2}+p^{2}\left(5\mu^{2}-m_{d}^{2}\right)}{3\mu^{2}-m_{d}^{2}}},\\
\omega_{3-}&\backsimeq& \sqrt{\frac{\mu^{2}-m_{d}^{2}}{3\mu^{2}-m_{d}^{2}}}p.\label{ss4}\
\end{eqnarray}
From the above relations it is noted that the excitation
$\omega_{1,2+}$ and $\omega_{3 +}$ have gaps identical to $2\mu$ and
$\sqrt{6\mu^{2}-2m_{d}^{2}}$, respectively.  The Nambu-Goldstone (NG)
bosons can be identified with excitations which are gapless as their
momentum goes to zero. Therefore, there are only $3$ gapless NG bosons
in the spectrum.  Due to symmetry consideration one may expect $5$ NG
bosons. If the conventional counting of NG bosons would apply, the
number of NG bosons $N$ is equal to the number of broken
generators. However, as has been previously noticed in
Lorentz-noninvariant systems the number of NG-modes can be less than
the number of broken generators \cite{shv}. A more accurate counting
role is due to the Nielsen-Chadha theorem \cite{nc} which states that
if $n_{1}$ and $n_{2}$ are the number of gapless excitations with
infrared dispersion laws $\omega_{1}\backsimeq p$ and
$\omega_{2}\backsimeq p^{2}$, respectively, then the number of NG
bosons $N\leq n_{1}+2n_{2}$. Therefore, there is no contradiction
since two of three gapless modes Eqs.~(\ref{ss1},\ref{ss4}) have
quadratic dispersion laws for long wavelengths. It is interesting to
note that at the BEC phase transition point when $\mu
\to m_{d}$ we have $\omega_{3-}\backsimeq\frac{p^{2}}{2\mu}$, and
all three gapless excitations become degenerate with the quadratic
dispersion law. Therefore, based on the naive continuity of spectrum
on the boundary of phase transition, one might have expected less
 NG bosons. One should also note that since we have gapless
mode with quadratic dispersion laws rather linear, the Landau
criterion \cite{lc} for superfluidity fails\footnote{According to the
Landau criterion, a fluid moving with group velocity $v$ is superfluid
if $v<v_{c}=\text{Min}(\frac{\omega_{i}(p)}{p})$ where the minimum is
taken with respect to all excitation modes and the
momentum $p$. If system has a quadratic dispersion of the low-energy
excitation modes, then $v_{c}=0$, and Landau criterion can not be
satisfied.}. Dissipationless flow in homogeneous bosonic system is one
of the manifestation of the interplaying  microscopic interaction
and the Bose-Einstein condensation. The question whether a
Bose-Einstein condensed  diquark gas in a more realistic model
really displays the Landau criterion for superfluidity remains to be
answered elsewhere.

The thermodynamical potential $\Omega_{d}(\mu, m_{d})$ at diquark
chemical potential $\mu$ is obtained through the partition function
$\mathcal{Z}$
\begin{equation}
\Omega_{d}(\mu, m_{d})=i\frac{\ln\mathcal{Z}}{V},  \hspace{2cm} \mathcal{Z}=\int \mathcal{D}\Phi^{\dag}\Phi e^{i\int d^{4}x \mathcal{L}_{d}},
\end{equation}
where $V$ denotes the spacetime volume of the system. The one-loop
contribution to the effective potential evaluated at the classical
minimum Eq.~(\ref{minp}) is obtained as 
\begin{eqnarray}
\Omega_{d}(\mu, m_{d})&=&\Omega^{MF}_{d}
-\frac{1}{2}i\int
\frac{d\omega}{2\pi}\int\frac{d^{3}p}{(2\pi)^{3}}\left[2\ln \Det
i\mathcal{D}_{1}^{-1}(\omega,p)+\ln \Det i\mathcal{D}_{3}^{-1}(\omega,p)\right],\nonumber\\
&=&-\frac{1}{4g}(\mu^{2}-m_{d}^{2})^{2}+\frac{1}{2}\int\frac{d^{3}p}{(2\pi)^{3}}\left[2\left(\omega_{1+}+\omega_{1-}\right)+\omega_{3+}+\omega_{3-}\right], \label{di-energy}\nonumber\\
\end{eqnarray}
 where $\mathcal{D}_{1,3}^{-1}$ is defined in Eq.~(\ref{prod}) and
 evaluated at the minimum of the classical action. The
 $\omega_{1,3\pm}$ is defined in Eq.~(\ref{ds1}). In the second line
 in the above equation, we have performed the integration over energy
 using contour integration. The first term $\Omega^{MF}_{d}$ in
 Eq.~(\ref{di-energy}) is the mean-field thermodynamical potential
 evaluated at the minimum of the classical action. The momentum
 loop-integrals in the diquark effective potential
 Eq.~(\ref{di-energy}) are divergent and have to be regularized. For
 simplicity we use a sharp cutoff $\Lambda_{D}$ in three-dimensional
 momentum space. Note that scalar field theory given in Lagrangian
 Eq.~(\ref{ac0}) is renormalizable (when higher order interaction
 terms are ignored), therefore any cutoff dependence can be removed by
 some kind of renormalization scheme. Here, we assume that the model
 has a natural ultraviolet cutoff which is defined for the validity of
 the model at high-energy.  We will investigate the implication of
 various choices for the cutoff.

One can show that the zeroth and first-order derivatives of the
thermodynamical potential Eq.~(\ref{di-energy}) with respect to
chemical potential $\mu$ at $\mu=m_{d}$ are continuous, but the
second-order derivative at $\mu=m_{d}$ is discontinuous. This
indicates a second-order phase transition between normal and diquark
condensate phases.  The charge density $\rho$ and the speed of sound
$v_{s}$ can be obtained from the thermodynamical potential
\begin{eqnarray}
\rho(\mu)&=&-\frac{\partial \Omega_{d}}{\partial \mu}, \label{mur}\\
v^{2}_{s}&=&\frac{\rho}{\mu}\frac{\partial \mu}{\partial \rho}. \
\end{eqnarray}
Using the above relation, the speed of sound at mean-field
approximation is obtained from the first term in the thermodynamical potential
Eq.~(\ref{di-energy}), 
\begin{equation}
v^{2}_{s}=\sqrt{\frac{\mu^{2}-m_{d}^{2}}{3\mu^{2}-m_{d}^{2}}}.
\end{equation}
Note also that this is identical to the coefficient of $\omega_{3-}$
in Eq.~(\ref{ss4}). At very high density $\mu\gg m_{d}$ we recover the
standard relativistic result $v_{s}=\frac{1}{\sqrt{3}}$ independent of
the model parameters. This indicates that at very high density the NG
boson propagates as  in a relativistic medium.

\section{Diquark BEC and the nuclear matter equation of state}
We incorporate the deconfinement effect by allowing that the active degrees
of freedom change from baryons to diquarks at a critical baryonic
critical chemical potential $\mu^{\star}_{b}$. For description of the
hadronic phase, we use the chiral $\sigma-\omega$ model with exact
global $SU(2)\times SU(2)$ symmetry, which contains a pseudoscalar
coupling between pions and massive isoscalar vector field
$\omega^{\mu}$ to nucleons,
\begin{eqnarray}
\mathcal{L}_{\text{h}}&=&\bar{\psi}_{n}[\gamma_{\mu}(i\partial^{\mu}-g_{vn}\omega^{\mu})-g_{n}\left(\sigma+i\gamma_{5}\vec{\tau}\vec{\pi}\right)
+\mu_{b}\gamma_{0}]\psi_{n}
-\frac{1}{4}F_{\mu\nu}F^{\mu\nu}\\
&+&\frac{1}{2}g_{v}^{2}\omega^{\mu}\omega_{\mu}\left(\sigma^{2}+\vec{\pi}^{2}\right)+\mathcal{L}_{\sigma\pi},\nonumber\\
\mathcal{L}_{\sigma\pi}&=&\frac{1}{2}\left(\partial_{\mu}\sigma\partial^{\mu}\sigma+\partial_{\mu}\vec{\pi}\partial^{\mu}\vec{\pi}\right)+\frac{m_{s}^{2}}{2}(\sigma^{2}+\vec{\pi}^{2})
+\frac{\lambda}{4}(\sigma^{2}+\vec{\pi}^{2})^{2}, \label{d0}\
\end{eqnarray}
where the field tensor is defined
$F_{\mu\nu}=\partial_{\mu}\omega_{\nu}-\partial_{\nu}\omega_{\mu}$ and
$\mu_{b}$ denotes the baryonic chemical potential. The nucleon mass
$M_{N}$ and the vector-meson mass $m_{v}$ at rest are generated
through spontaneous symmetry breaking by the $\sigma$-field
\begin{equation}
M_{N}(\sigma)=g_{n}\sigma, \hspace{2cm} m_{v}(\sigma)=g_{v}\sigma.
\end{equation}
The energy density for the hadronic phase can be obtained from
Eq.~(\ref{d0}) in the conventional mean-field approximation. Having
used the equation of motion for the $\omega^{0}$ field, we obtain the energy
density as
\begin{eqnarray}
\varepsilon_{h}=V(\bar{\sigma})+\frac{g_{vn}^{2}\rho_{b}^{2}}{2m^{2}_{v}(\bar{\sigma})}
&+&\frac{2}{\pi^{2}}\int^{k_{F}}_{0} dk~k^{2}\omega_{N}(k,\bar{\sigma})\nonumber\\
&-&\frac{2}{\pi^{2}}\int^{\Lambda_{N}}_{0} dk~k^{2}
\left(\omega_{N}(k,\bar{\sigma})-\omega_{N}(k,\bar{\sigma}_{0})\right),\label{energy1}\\
V(\bar{\sigma})=\frac{m_{s}^{2}}{2}\left(\bar{\sigma}^{2}-\bar{\sigma}^{2}_{0}\right)&+&\frac{\lambda}{4}\left(\bar{\sigma}^{4}-\bar{\sigma}^{4}_{0}\right)\label{energy}\
\end{eqnarray}
where we defined
$\omega_{N}(k,\bar{\sigma})=\sqrt{k^{2}+M_{N}^{2}(\bar{\sigma})}$ and 
$\bar{\sigma}$ ( and $\bar{\sigma}_{0}$) denotes the mean-value of
the sigma field in the nuclear matter medium (and in the vacuum). We have
also included the corresponding Dirac-sea of the nucleon up to a
ultraviolet cutoff $\Lambda_{N}$. In principle, in renormalizable
models such as the linear sigma model, divergences can be absorbed
into the coefficient of interaction terms by employing a
renormalization scheme. However, it has been shown that for this kind
of models with heavy fields, such terms give rise to
unnaturalness\footnote{Furnstahl {\em et. al.} \cite{fs} have shown
that the one-baryon-loop vacuum contribution in renormalized models like
the linear sigma model and the Walecka model gives rise to large {\em
unnatural} coefficients based on the ``naive dimensional analysis''
proposed by Georgi and Manohar
\cite{la2}.} indicating that quantum vaccum is not adequately
described by long-range degrees of freedom \cite{fs}.  
We have recently shown that nuclear matter properties can be better
reproduced if one treats the vacuum loop explicitly \cite{me0}. We
discard the short-range physics which cannot be described by
long-range degrees of freedom by the nucleonic ultraviolet cutoff
$\Lambda_{N}$. The baryonic density $\rho_{b}$ is related to the Fermi
momentum $k_{F}$ by
\begin{equation}
\rho_{b}=\frac{2}{\pi^{2}}\int^{k_{F}}_{0}dk k^{2}=\frac{2k^{3}_{F}}{3\pi^{2}}.
\end{equation}
At any density $\rho_{b}$, $\varepsilon_{h}$ is stationary with respect
to the mean-value of the scalar field. Therefore, the scalar
mean-field $\bar{\sigma}$ is obtained via the self-consistency conditions:
\begin{equation}
\frac{d\varepsilon_{h}}{d\bar{\sigma}}=0. \label{sel0}\ 
\end{equation}
This equation is numerically solved for every point of density. The
baryonic chemical potential $\mu_{b}$ can be obtained through the Hugenholtz-van
Hove theorem via the energy density
\begin{equation}
\mu_{b}=\frac{\partial \varepsilon_{h}}{\partial \rho}=\frac{g_{vn}^{2}\rho_{b}}{m^{2}_{v}(\bar{\sigma})}+\omega_{N}(k_{F}, \bar{\sigma}),
\end{equation}
where the value of $\bar{\sigma}$ for a given density is obtained
via Eq.~(\ref{sel0}). Then the thermodynamical potential $\Omega_{\text{h}}$ and the hadronic pressure $P_{\text{h}}$ are
obtained through a Legendre transformation and can be written as
\begin{eqnarray}
\Omega_{\text{h}}&=&-P_{\text{h}}=\varepsilon_{h}-\mu_{b}\rho_{b}=V(\bar{\sigma})-\frac{g_{vn}^{2}\rho_{b}^{2}}{2m^{2}_{v}(\bar{\sigma})}\nonumber\\
&&+4\int\frac{d^{3}k}{(2\pi)^{3}}\left(\omega_{N}(k, \bar{\sigma})+\frac{g_{vn}^{2}\rho_{b}}{m^{2}_{v}(\bar{\sigma})}-\mu_{b}\right)\Theta\left(\mu_{b}-\omega_{N}(k, \bar{\sigma})-\frac{g_{vn}^{2}\rho_{b}}{m^{2}_{v}(\bar{\sigma})}\right)
\nonumber\\
&&-\frac{2}{\pi^{2}}\int^{\Lambda_{N}}_{0} dk~k^{2}
\left(\omega_{N}(k,\bar{\sigma})-\omega_{N}(k,\bar{\sigma}_{0})\right)
.\label{nu} \
\end{eqnarray}
 For the deconfined phase we consider matter made of diquarks coupled
 to $\sigma$ and $\vec{\pi}$ and we use the diquark Lagrangian introduced in the previous section. The new input here
 is that we couple the diquark field to the scalar field in order to
 generate the diquark mass dynamically via spontaneous symmetry
 breaking. 
\begin{equation}
\mathcal{L}_{\text{q}}=
(\partial_{0}+i\mu)\Phi^{\dag}(\partial_{0}-i\mu)\Phi-\left(\partial_{i}\Phi^{\dag}\right)\left(\partial_{i}\Phi\right)
-(g_{d}\sigma)^{2}\Phi^{\dag}\Phi-g\left(\Phi^{\dag}\Phi\right)^{2}+ \mathcal{L}_{\sigma\pi}. \label{lq}
\end{equation}
The Lagrangian $\mathcal{L}_{\sigma\pi}$ is defined in Eq.~(\ref{d0}).
We incorporate the diquark BEC effect in the fashion presented in the
previous section, namely by employing the diquark field
parametrization Eq.~(\ref{dw}).  We treat the diquark fields at the
one-loop level. Therefore, the thermodynamical potential
$\Omega_{\text{q}}$ and deconfined pressure $P_{\text{q}}$ are given
by
\begin{equation}
\Omega_{\text{q}}=-P_{\text{q}}=\Omega_{d}(\mu,m_{d}(\bar{\sigma}))+V(\bar{\sigma})-\frac{3}{2\pi^{2}}\int_{0}^{\Lambda_{D}} k^{2} dk\sqrt{k^{2}+m_{d}^{2}(\bar{\sigma}_{0})},\label{di-ter}
\end{equation}
where the diquark mass is related to the $\sigma$-field: 
\begin{equation}
m_{d}(\sigma)=g_{d}\sigma.\label{di-mas}
\end{equation}
The $\Omega_{d}$ and $V(\bar{\sigma})$ in Eq.~(\ref{di-ter}) are
defined in Eqs.(\ref{di-energy}) and (\ref{energy}), respectively. The
last term in Eq.~(\ref{di-ter}) is the constant diquark contribution
at zero density which shifts the minimum of the vacuum thermodynamical
potential to zero.  Note that the diquark mass now dynamically changes
in medium through the in-medium $\sigma$-field.  This does not alter
the possible formation of the diquark BEC as far as $\mu^{2}>m^{2}_{d}$
where $\mu$ is the diquark chemical potential. In contrast to the Bag
model and similarly to the NJL model, our diquark model dynamically
generates a density-dependent bag constant and density dependent
effective constituent diquark masses.  With the diquark
thermodynamical potential Eq.~(\ref{di-ter}) and the density
Eq.~(\ref{mur}) one can construct the energy density through the
Legendre transformation,
\begin{eqnarray}
\varepsilon_{q}=\Omega_{\text{q}}+\mu\rho(\mu)&=&V(\bar{\sigma})+\frac{1}{4g}\left(\mu^{2}-m^{2}_{d}(\bar{\sigma})\right)\left(3\mu^2+m^{2}_{d}(\bar{\sigma})\right)\nonumber\\
&+&\frac{1}{\pi^{2}}\int_{0}^{\Lambda_{D}}\frac{p^{4}dp}{\sqrt{p^{2}+\mu^{2}}}\nonumber\\
&+&\frac{1}{2\pi^{2}}\int_{0}^{\Lambda_{D}}p^{2}dp\left(\frac{\left(p^{2}-m^{2}_{d}(\bar{\sigma})\right)\omega_{3+}\omega_{3-}+\omega^{2}_{3+}\omega^{2}_{3-}-p^{2}\mu^{2}}{\omega_{3+}\omega_{3-}(\omega_{3+}+\omega_{3-})}\right)\nonumber\\
&-&\frac{3}{2\pi^{2}}\int_{0}^{\Lambda_{D}}k^{2} dk \sqrt{k^{2}+m_{d}^{2}(\bar{\sigma}_{0})},\label{energy2}\
\end{eqnarray}
where for the right hand side we have made use of the
definition of $\rho(\mu)$ and $\Omega_{\text{q}}$ given in
Eqs.~(\ref{mur},\ref{di-ter}). The $\omega_{3\pm}$ and $V(\bar{\sigma})$ are defined
in Eqs.~(\ref{ds1}) and (\ref{energy}), respectively. Using
Eq.~(\ref{ds1}) we obtain the following relations which facilitate the
computation
\begin{eqnarray}
\omega_{3+}\omega_{3-}&=&p\sqrt{p^{2}+2(\mu^{2}-m^{2}_{d}(\bar{\sigma}))},\nonumber\\
\omega_{3+}+\omega_{3-}&=&\sqrt{2}\sqrt{p^{2}+3\mu^{2}-m^{2}_{d}(\bar{\sigma})+p\sqrt{p^{2}+2(\mu^{2}-m^{2}_{d}(\bar{\sigma}))}}.\
\end{eqnarray}
The in-medium mean-value of the scalar field and the corresponding
diquark mass is obtained similarly to the hadronic sector by minimization
\begin{equation}
\frac{d\varepsilon_{q}}{d\bar{\sigma}}=0.  
\end{equation}

The diquark chemical potential and density ($\mu$,$\rho$) are related
to baryonic ones ($\mu_{b}$, $\rho_{b}$) by: $\mu=\frac{2}{3}\mu_{b}$
and $\rho=\frac{3}{2}\rho_{b}$. The binding energy per baryon is
defined as 
\begin{equation}
\frac{E_{B}}{A}=\frac{\varepsilon_{h,q}}{\rho_{b}}-M_{N}(\bar{\sigma}_{0})=-\frac{P_{h,q}}{\rho_{b}}+\mu_{b}-M_{N}(\bar{\sigma}_{0}), 
\end{equation}
where $\varepsilon_{h,q}$ and $p_{h,q}$ are defined in
Eqs.~(\ref{energy1},\ref{energy2}) and Eqs.~(\ref{nu},\ref{di-ter}), respectively.

\section{Nuclear-diquark phase transition at high density}
 We incorporate the deconfinement effect by demanding that as we
 increase the density in the hadronic phase at a critical Fermi
 momentum $k^{\star}_{F}$ corresponding to the baryonic chemical potential
 $\mu_{b}^{\star}$, the baryonic degrees of freedom are replaced by
 diquark ones while keeping the description of the auxiliary scalar
 fields intact above the transition. We do not couple the vector-meson
 field to diquarks.  The exact value of the deconfinement chemical
 potential $\mu_{b}^{\star}$ is not known. Therefore, we will consider
 the implication of different choices for $\mu_{b}^{\star}$ and
 $k^{\star}_{F}$. 

  In the hadronic phase, our model contains six unknown parameters
  $\Lambda_{N}$, $m^{2}_{s}$, $\lambda$, $g_{n}$, $g_{vn}$ and
  $g_{v}$. We take the hadronic ultraviolet cutoff $ \Lambda_{N}=324$
  MeV \cite{me0}. In this model the acceptable range of the ultraviolet cutoff $
  \Lambda_{N}$ is rather low \cite{me0}. This empirical fact has been
  also observed by other authors within the NJL model with nucleonic
  degrees of freedom \cite{em1}. This is in contrast with the case
  that the model is defined in terms of quark degrees of freedom where
  the allowed range of the cutoff is typically larger. Due to
  spontaneous symmetry breaking the scalar mean field acquires a
  nonzero vacuum expectation value which is equal to the pion
  decay. The pions are then the massless Nambu-Goldstone bosons.  For
  the given $ \Lambda_{N}=324$ MeV, one can determine the scalar
  potential couplings $m^{2}_{s}$ and $\lambda$ so that the mean-value
  of the scalar field becomes identical to the empirical pion decay at
  vacuum $\bar{\sigma}_{0}=93$ MeV. We take the empirical nucleon mass
  in vacuum $M_{N}(\bar{\sigma}_{0})=939$ MeV and vector-meson mass in
  vacuum $m_{v}(\bar{\sigma}_{0})=783 $ MeV. These choices fix the
  coupling $g_{n}=M_{N}/f_{\pi}=10.1$ and
  $g_{v}=m_{v}/f_{\pi}=8.42$. The vector coupling $g_{vn}$ is treated
  as a free parameter and is adjusted in the medium in such a way that
  the EoS reproduces nuclear-matter saturation properties, namely a
  binding energy per nucleon $E_{B}/A=-15.75$ MeV at a density
  corresponding to a Fermi momentum of
  $k_{F}=256~\text{MeV}=1.3~\text{fm}^{-1}$ \cite{wal}. In this way
  all parameters in the confined phase are fixed uniquely (see set NM
  in table 1).
\begin{table}
\centering
\begin{tabular}{lllllll}
\hline
\hline
parameter&$\Lambda_{N}(\text{MeV})$  &$m^{2}_{s}(\text{GeV}^{2})$ & $\lambda$ & $g_{vn}$  & $g_{n}$ &$g_{v}$ \\
\hline
set NM&324&+0.026&25&6.54&10.1&8.42\\
\hline
\hline
\end{tabular}
\caption{The value of various couplings in hadronic phase for set NM is given. This parameter set reproduces the empirical saturation point
($E_{b}/A=-15.75 ~\text{MeV}, \rho_{0}=0.15~\text{fm}^{3}$).}
\end{table}

The sigma mass can be obtained from
\begin{equation}
m^{2}_{\sigma}=\frac{\partial^{2}\Omega_{h}}{\partial\bar{\sigma}^{2}}, \label{smm}
\end{equation} 
where the derivatives are evaluated at the scalar mean-field
$\bar{\sigma}$ solution of the gap equation. The corresponding sigma
meson mass $m_{\sigma}$ in vaccum for the above parameters (set NM in
table 1) is $810$ MeV. 
In order to estimate the stiffness of the
nuclear matter EoS, one may compute the compression modulus at the saturation density. It is defined as
\begin{equation}
K=-9\frac{d\Omega_{h}}{d\rho_{b}}=9\rho_{b}\frac{\partial^{2}\varepsilon_{h}}{\partial \rho_{b}^{2}}. \label{comp}
\end{equation}

The compression modulus at saturation density for the parameter set NM is
$K=455$ MeV. Note that in the original $\sigma-\omega$ model of Boguta
the compression modulus of nuclear matter is unacceptably large about
$K=650$ MeV
\cite{bog}. Here, due to the inclusion of the Dirac-sea we are
relatively able to reduce the compression modulus (see also
Ref.~\cite{me0}). This value is still bigger than the empirical one
which is believed to be about $K=200-300$ MeV
\cite{ems,ems1}. Nevertheless, here we use only the nuclear matter compression modulus as a relative 
measure of stiffness of the deconfined EoS. It has been very difficult
to overcome the unphysically high values of compressibility in the
$\sigma-\omega$ model and in the Walecka model. In the next section we
will introduce a non-conventional scenario in which the nuclear matter
compressibility can be lowered.
\begin{table}
\centering
\begin{tabular}{lllll}
\hline
\hline
Parameter ($\mu^{\star}_{b}=1214.5$ MeV) &set A1 & set A2 &set A3 & set A4\\
\hline
$g$ &14.1  & 13.47& 12.73&11.23\\ 
\hline
$g_{d}$ & 6.85& 6.7& 6.5&6.0\\
$m_{d}(\rho=0)$(MeV)& 637.1&623.1& 604.5&558.0\\
\hline
\hline
$\rho^{h}_{b}/\rho_{0}$ &3.0&3.0&3.0&3.0\\ 
$\rho^{d}_{b}/\rho_{0}$ &3.0&4.0&5.4&8.6\\ 
\hline
\hline
\end{tabular}
\caption{The parameters $g_{d}$ and $g$ in the deconfined phase for the sets A1-A4 for the critical deconfinement chemical 
potential $\mu^{\star}_{b}=1214.5$ MeV. We assume as
ultraviolet cutoff for all parameter sets $\Lambda_{D}=900$ MeV. The
baryonic density at the transition point in hadronic $\rho_{b}^{h}$
and diquark $\rho^{d}_{b}$ parts in units of the normal nuclear matter
density $\rho_{0}=0.15~\text{fm}^{3}$ are also given. }
\end{table}

For zero temperature, two phases of hadrons and diquarks can be in equilibrium if their
chemical potentials and pressures are equal (Gibbs'
conditions):
\begin{equation} 
P_{q}(\mu_{b}^{\star})=P_{h}(\mu_{b}^{\star}),
 \hspace{2cm} \frac{3}{2}\mu=\mu_{b}=\mu_{b}^{\star}, \label{gib}
\end{equation}
where $P_{h}$ and $P_{q}$ are defined in Eqs.~(\ref{nu}) and
(\ref{di-ter}), respectively. First, we assume that the transition to
the diquark BEC matter occurs at the baryonic chemical potential
$\mu^{\star}_{b}=1214.5$ MeV which corresponds to the baryonic density
$\rho^{h}_{b}=3\rho_{0}$ from the hadronic side.
For the deconfined phase, we have three
unknown parameters: the ultraviolet cutoff $\Lambda_{D}$, the
diquark-diquark coupling $g$ and the diquark-scalar field coupling
$g_{d}$. We assume the deconfined ultraviolet cutoff to be
$\Lambda_{D}=900$ MeV and fix the other two unknown parameters by
Gibbs' conditions Eq.~(\ref{gib}). We require that at the phase
transition the baryonic chemical potential of both hadronic and
diquark parts to be equal
$\mu_{b}=\frac{3}{2}\mu=\mu_{b}^{\star}=1214.5$ MeV.  Having obtained
the nuclear matter EoS, one can immediately read off the
hadronic pressure at the critical deconfinement chemical potential:
$P_{h}(\mu^{\star}_{b}=1214.5~\text{MeV})=89.8~\text{MeV}/\text{fm}^{3}$. Therefore,
the Gibb's condition for equilibrium of two phases is satisfied if
$P_{h}(\mu^{\star}_{b})=P_{q}(\mu^{\star}_{b})=89.8~\text{MeV}/\text{fm}^{3}$.
This gives a non-trivial constraint on the diquark-model parameters. There
exist many solutions which satisfy the above requirements, each of
them corresponds to a different BEC condensate. The values of
parameters $g$ and $g_{d}$ for four sets A1-A4 with a fixed
$\Lambda_{D}=900$ MeV are given in table 2. We have also given in
table 2 the value of the diquark mass in vacuum using
Eq.~(\ref{di-mas}). Note that the values of the scalar diquark mass in
vacuum for various parameter sets given in table 2 is within the
acceptable range which has been estimated by many authors in order to
describe the baryon properties in vacuum \cite{n-d,dm0,dm}.
\begin{figure}[!tp]
       \centerline{\includegraphics[width=8.5 cm] {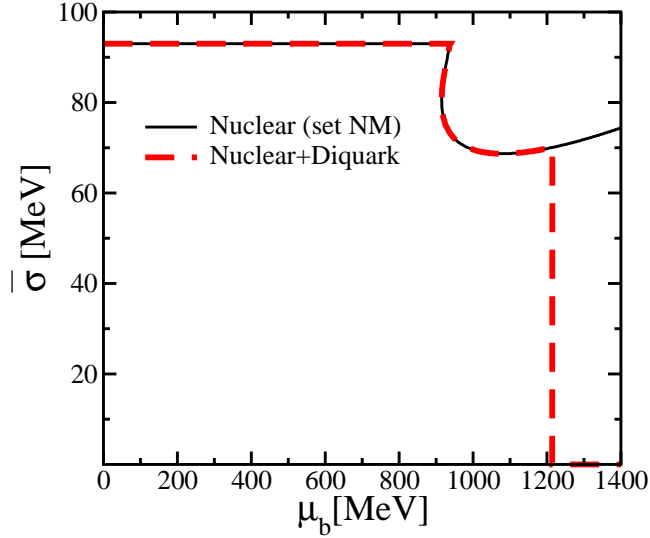}}
       \caption{In-medium pion decay constant $\bar{\sigma}$ as a
       function of baryonic chemical potential $\mu_{b}$. We use the
       parameter set NM given in table 1 for nuclear matter and
       parameter sets $A1-A4$ given in table 2 for diquark
       matter.\label{pm0}}
\end{figure}

In Fig.~\ref{pm0}, we show the mean-value of the scalar field $\bar{\sigma}$ with
respect to the baryonic chemical potential $\mu_{b}$.  The solid line
denotes the mean-value of the scalar field in the nuclear matter medium
when the deconfinement to diquark matter is not incorporated.
For all diquark parameter sets (given in table 2) chiral phase
transition takes place at the critical deconfinement chemical potential
$\mu_{b}^{\star}=1214.5$ MeV. We will show later that this will also be
the case if the critical deconfinement chemical potential is taken
lower value.

\begin{figure}[!tp]
       \centerline{\includegraphics[width=15 cm] {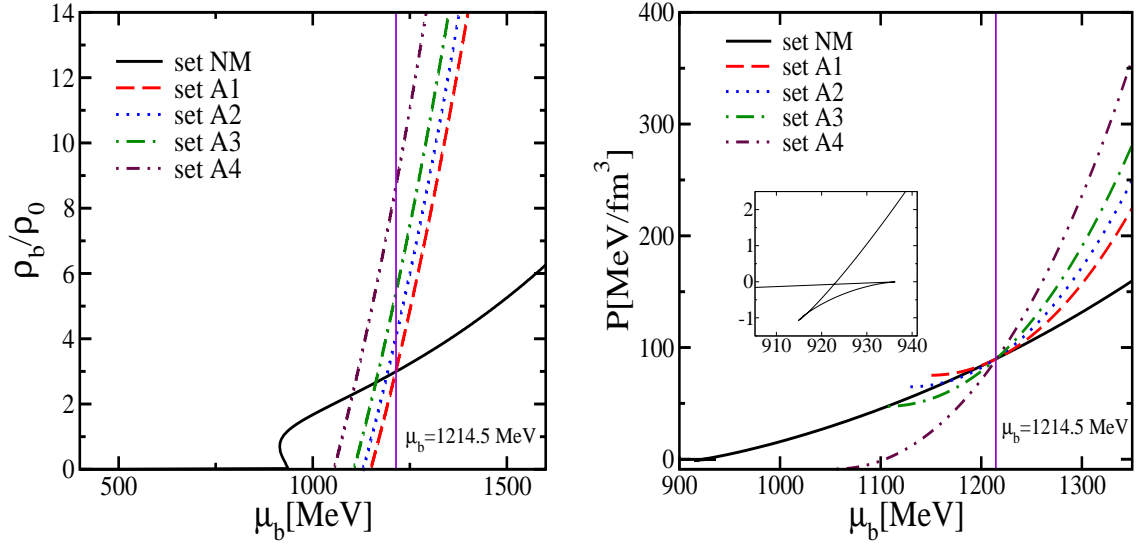}}
       \caption{Baryon number densities (left) in units of nuclear
       matter density $\rho_{0}=0.15 fm^{-3}$ and pressure (right) as
       functions of the baryonic chemical potential $\mu_{b}$ for
       various parameter sets given in tables 1 and 2. The critical
       deconfinement line $\mu^{\star}_{b}=1214.5$ MeV is also
       shown. The insert shows the region of the nuclear matter liquid-gas phase transition.\label{pr0}}
\end{figure}

In Fig.~\ref{pr0} (right pannel), we show the diquark pressure $P_{q}$
for sets A1-A4 (in table 2) and the hadronic pressure $P_{h}$ for set NM
(in table 1) as a function of the baryonic chemical potential
$\mu_{b}$. For comparison we also display the nuclear matter solution
for the parameter set NM when the transition to a diquark BEC is not
incorporated. It is obvious that the EoS becomes considerably softer
if deconfinement is allowed. Therefore, the system is energetically
favored to undergo phase transition.  By comparing different
diquark EoS obtained by using various parameter sets given in table 2,
it is observed that a bigger diquark-diquark coupling $g$ leads to a
stiffer EoS, whilst a bigger coupling $g$ leads to a smaller diquark
condensate, see Eq.~(\ref{minp}). Therefore, the bigger the diquark
BEC condensate the softer the EoS will be. The inserted plot in
Fig.~\ref{pr0} shows the first-order liquid-gas phase transition of
nuclear matter which takes place at $\mu_{b}=923.25$ MeV. The
first-order phase transition is manifested by the appearance of several
branches of $P_{h}(\mu_{b})$. At a fixed baryonic chemical potential,
only the highest pressure corresponds to a stable phase. The slope of
curves $P(\mu_{b})$ exhibits a jump in density. In Fig.~\ref{pr0}
(left pannel), we show the baryonic density $\rho_{b}$ in units of
nuclear matter density $\rho_{0}$ as a function of the baryonic
chemical potential, we have also plotted the critical deconfinement
chemical potential $\mu^{\star}_{b}$. In general, there are two
separated density jumps with respect to the chemical potential: one
corresponds to the standard nuclear matter liquid-gas phase transition
and the second one corresponds to the deconfinement phase transition. The
value of baryonic density below and above deconfinement is given in
table 2. Depending on model parameters the jump at the deconfinement
transition can change.  The softest EoS set A4 results in a high
baryonic density above the phase transition since the diquark BEC
condensate is bigger and consequently it can accommodate more baryon
density. Note also that the set A4 has the lowest diquark mass in
vaccum, see table 2. For the model parameter (set A1) with rather
heavy diquark in vaccum, transition to diquark matter takes place
without a density jump, i. e. the first derivative of the thermodynamical
potential with respect to the chemical potential is continuous. This
indicates a second-order phase transition between hadronic and diquark
matter. In this case, although deconfinement and chiral phase
transition coincides, their phase orders are different.

\begin{figure}[!tp]
       \centerline{\includegraphics[width=8 cm] {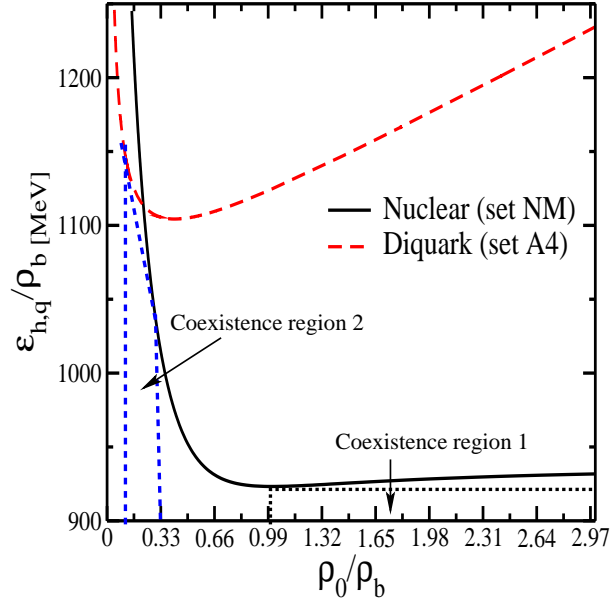}}
       \caption{Energy per baryon number as functions of inverse
       baryonic density in units of the nuclear matter density, 
       i. e. $(\rho_{b}/\rho_{0})^{-1}$ when the deconfinement
       chemical potential is taken $\mu^{\star}_{b}=1214.5$ MeV. The
       dotted lines represent the Maxwell-constructions which
       determine the regions of coexistence. The coexistence region 1
       contains the nucleonic gas and nuclear droplets and the
       coexistence region 2 contains mixed nuclear and diquark BEC
       matter.  \label{pmax2}}
\end{figure}

In order to consider the possible coexistence regions which emerge
from the first-order phase transition, we perform Maxwell
constructions. In Fig.~\ref{pmax2}, we show the energy per baryon
$\varepsilon_{h,q}/\rho_{b}$ as a function of inverse baryonic density
$(\rho_{b}/\rho_{0})^{-1}$ for the parameter set A4 given in table 2. It
exhibits a typical Van der Waals-like diagram. Using the following
thermodynamical consistency equation and Gibb's conditions,
\begin{equation}
\frac{\partial}{\partial \rho^{-1}_{b}}\left(\frac{\varepsilon_{h,q}}{\rho_{b}}\right)=-P_{h,q}, \label{maxcon}
\end{equation}
one can identify the coexistence region as a function of inverse
density. There exist two separated coexistence regions which are
denoted in Fig.~\ref{pmax2} as coexistence region $1$ and $2$. At low
density $0\leq\rho_{b}\leq\rho_{0}$ in the hadronic phase we have the
typical coexistence between nucleonic gas and nuclear droplet. At
rather high density $3\rho_{0}\leq\rho_{b}\leq\rho^{d}_{b}$ (for
values of $\rho^{d}_{b}$ see table 2), there exist a coexistence
region between nuclear matter and chirally restored diquark BEC
matter. At zero temperature, the coexistence region between the hadronic
and diquark BEC phase seems to be quite large in the density axis for a
weaker diquark-diquark interaction or a bigger diquark BEC (see
parameter set A4). Forcrand {\em et al.} \cite{la4} have recently
found some indication for the coexistence region of the hadron and the
plasma phases based on the canonical approach of lattice QCD (for
4-flavour and pion mass $m_{\pi}\approx 350$ MeV). Their extrapolation
indicates that the coexistence region might be quite wide in density
$0.50(5)~\text{B}/\text{fm}^{3}\lesssim\rho \lesssim
1.8(3)~\text{B}/\text{ fm}^{3}$. It is of great interest to see if 
this coexistence region will survive in the chiral limit using a fully realistic simulation of QCD.

\section{Nuclear-diquark phase transition at low density}
In this section, we consider the possibility that the deconfining
transition happens at a lower baryonic chemical potential. In
particular, we study the extreme case where the deconfinement phase
transition takes place right at the nuclear matter saturation Fermi
momentum, namely at $k_{F}^{\star}=256$ MeV which is approximately
equal to the QCD confinement scale $\Lambda_{\text{QCD}}$.  This
corresponds to a baryonic chemical potential
$\mu_{b}^{\star}=M_{N}(\bar{\sigma}_{0})-E_{B}/A=923.25$ MeV. At
the saturation point, the binding energy per nucleon is taken
$E_{B}/A=15.75$ MeV. Above the deconfinement phase transition the
vector-meson does not appear in the deconfined Lagrangian
$\mathcal{L}_{q}$ Eq.~(\ref{lq}) and one may be afraid that nuclear matter in
this model would collapse \cite{bog}. However, we will show
that the repulsive diquark interaction induces enough repulsion to
reproduce stable matter similar to conventional nuclear matter.
\begin{table}
\centering
\begin{tabular}{lllll}
\hline
\hline
Parameter ($\mu^{\star}_{b}=923.25$ MeV) &set B1 & set B2 &set B3 & set B4\\
\hline
$g$&18.16 &16.43  & 15.46 & 14.34\\ 
\hline
$g_{d}$ &4.8& 4.56& 4.40& 4.18 \\
$m_{d}(\rho=0)$(MeV)&446.4 &424.1 &409.2& 388.8\\
\hline
\hline
$\rho^{h}_{b}/\rho_{0}$&1.0 &1.0&1.0&1.0\\ 
$\rho^{d}_{b}/\rho_{0}$&0.23 &1.0&1.6& 2.24\\ 
\hline
\hline
\end{tabular}
\caption{The parameters $g_{d}$ and $g$ in the
deconfined phase for the sets B1-B4 for the critical deconfinement
chemical potential $\mu^{\star}_{b}=923.25$ MeV. The diquark
ultraviolet cutoff for all parameter sets is taken $\Lambda_{D}=650$
MeV.}
\end{table}

For the case of early deconfinement, a lower
diquark ultraviolet cutoff $\Lambda_{D}$ is allowed since we are interested in
physics of rather low baryonic chemical potential. We assume as diquark ultraviolet
cutoff $\Lambda=650$ MeV. Later we will investigate the effect of
choosing a higher value for the ultraviolet cutoff.  The acceptable range of
the diquark-scalar field coupling $g_{d}$ and the diquark-diquark coupling
$g$ is determined via the Gibb's condition.  Since from the hadronic
part at the saturation point we have that the hadronic pressure $P_{h}(\mu_{b}^{\star})=0$,
we require that also the diquark pressure $P_{q}(\mu_{b}^{\star}=923.25
\text{MeV})=0$. In table 3 we give the values of the couplings in four parameter sets
B1-B4 obtained by the above-mentioned procedure.

\begin{figure}[!tp]
       \centerline{\includegraphics[width=15cm] {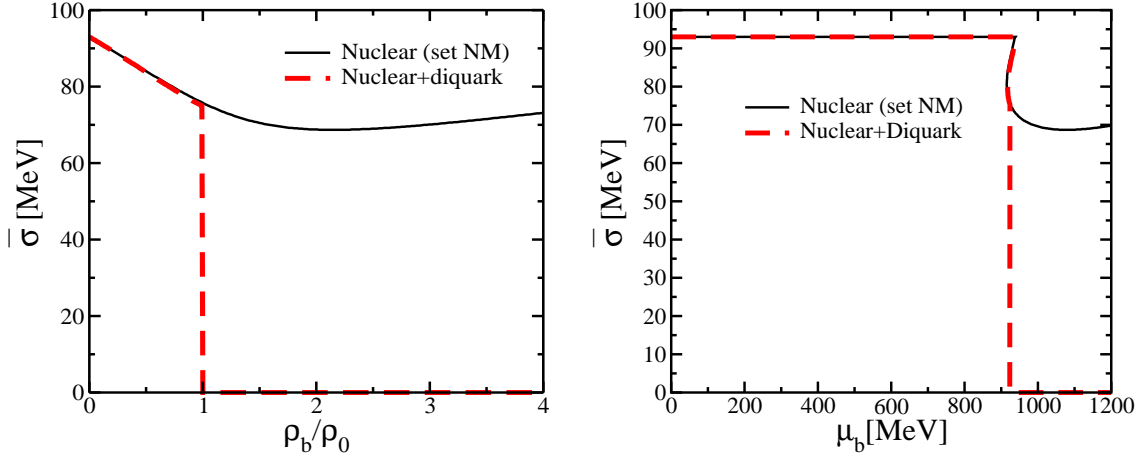}}
       \caption{In-medium pion decay constant $\bar{\sigma}$ as functions of
       density $\rho_{b}/\rho_{0}$ (left) in unit of nuclear matter
       density $\rho_{0}=0.15 fm^{-3}$ and baryonic chemical potential
       $\mu_{b}$ (right). We have used parameter sets NM and set B2 given in tables
       1 and 3.\label{pm1}}
\end{figure}

In Fig.~\ref{pm1}, we show the in-medium pion decay constant
$f_{\pi}=\bar{\sigma}$ as a function of the baryonic chemical potential
$\mu_{b}$ (left pannel) and baryonic density $\rho_{b}/\rho_{0}$
(right pannel). The chiral restoration for all parameter sets given in
table 3 is first order and takes place right at the critical
deconfinement chemical potential $\mu^{\star}_{b}$.  The solid line
denotes the mean-value of the scalar field in the nuclear matter medium
without deconfinement effect.  It is well known that in the
$\sigma-\omega$ model with dynamical $\omega$-mass, chiral restoration
does not occur since $\bar{\sigma}$ tends to grow at high
density. This can also be seen in Figs.~\ref{pm0} and \ref{pm1}, where
$\bar{\sigma}$ at higher chemical potential and density in nuclear
medium (solid line) bends upward. In fact this feature of the model is
closely connected to the stability of the system. The sigma field
plays the role of the chiral partner of the pion and at the same time
it is the mediator of medium-range nucleon-nucleon attraction and via
the $\omega$-mass also the short range repulsive is related to the
sigma field. Therefore, as its mass becomes smaller the attraction
between nucleons becomes stronger which may destroy the stability of
nuclear matter. This effect is not present in non-chiral type models
such as the Walecka model
\cite{wal}. However, we will show that in our approach the chiral
restoration does not obstruct the stability of the system. 
\begin{figure}[!tp]
       \centerline{\includegraphics[width=9 cm] {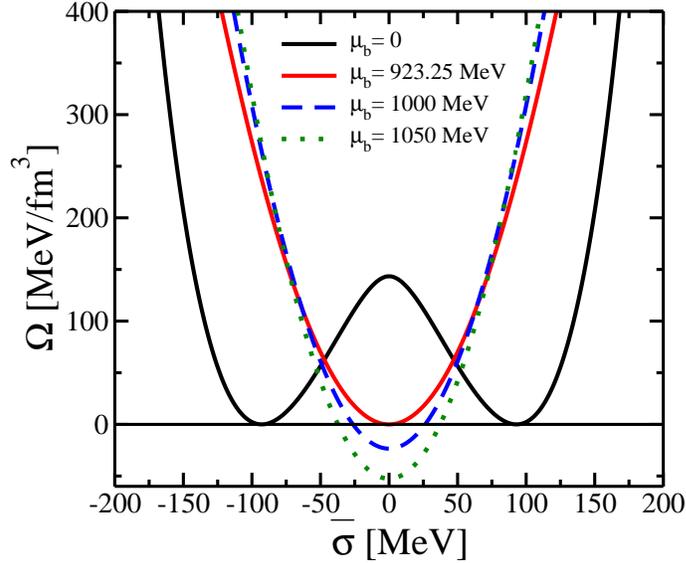}}
       \caption{Thermodynamic potential $\Omega_{q}$(and $\Omega_{h}$)
       as a function of auxiliary variable $\bar{\sigma}$ (pion decay)
       for parameter set NM (for $\mu_{b}=0$) and for other values of
       $\mu_{b}$ the parameter set B2 is used. 
         \label{pp} }
\end{figure}

 The thermodynamic potentials $\Omega_{q}$ and $\Omega_{h}$ as a
 function of the auxiliary variable $\bar{\sigma}$ are shown in
 Fig.~\ref{pp}. In vacuum the minimum of the thermodynamic potential
 $\Omega_{h}$ is located at the pion decay $\bar{\sigma}=93$ MeV. As
 we increase the chemical potential right at the critical
 deconfinement chemical potential $\mu_{b}^{\star}$ where we switch to
 the diquark thermodynamic potential $\Omega_{q}$, there is a strong first
 order phase transition from the vacuum to the chirally restored phase
 with $\bar{\sigma}=0$. Once the phase transition has taken place, the
 minimum of $\Omega_{q}$ at $\bar{\sigma}=0$ remains with increasing
 chemical potential, but the lowest $\Omega_{q}$ configuration becomes
 more and more negative $\Omega_{q}<0$, indicating an increase in
 pressure $P_{q}>0$ with density. At the critical chemical potential
 $\mu^{\star}_{b}=923.25$ MeV, both minima of $\Omega_{h}$ and
 $\Omega_{q}$ have zero pressure. One represents the non-perturbative
 broken vacuum at density zero and the other represents the chirally
 restored diquark BEC phase at non-zero density $\rho_{b}$. Note that
 we construct the deconfinement phase transition in such a way that
 the chiral phase transition results.  The chiral restoration is first
 order and coincides with the deconfinement critical chemical
 potential.

\begin{figure}[!tp]
       \centerline{\includegraphics[width=15 cm] {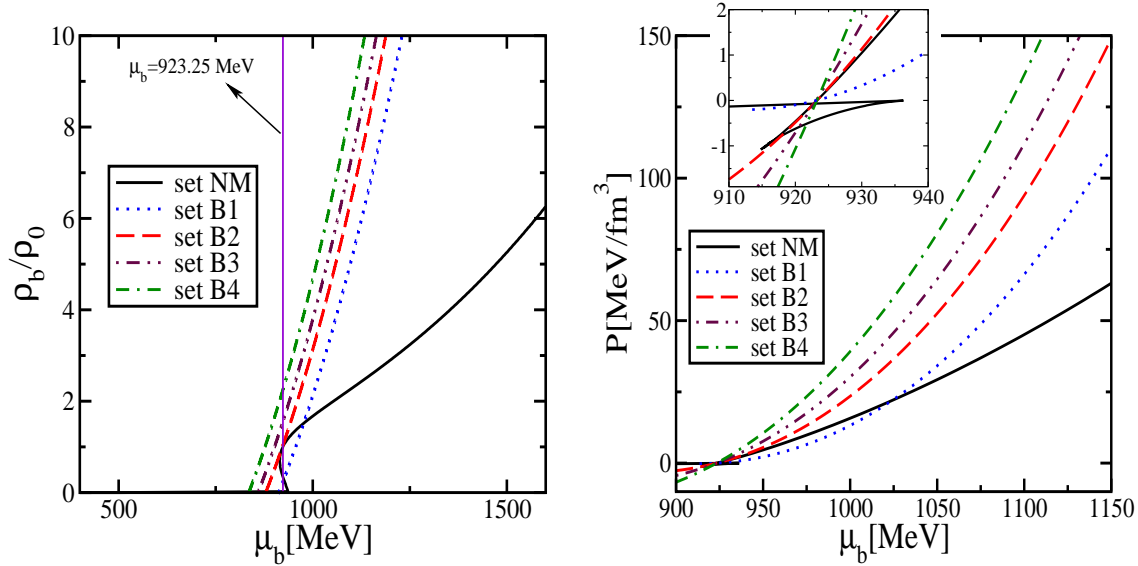}}
       \caption{Baryon number densities (left) in units of nuclear
       matter density $\rho_{0}=0.15 fm^{-3}$ and pressure (right) as
       functions of the baryonic chemical potential $\mu_{b}$ for
       various parameter sets given in tables 1 and 3. For comparison
       we have also plotted the nuclear matter solution (solid line)
       for the parameter set NM when transition to diquark matter is not
       taken into account.\label{pr1}}
\end{figure}

In Fig.~\ref{pr1} (right pannel), we show the nuclear matter pressure
$P_{h}$ for set NM (in table 1) and the diquark matter pressure
$P_{q}$ for set B1-B4 (in table 3) as a function of the baryonic
chemical potential. It is seen that above the critical deconfinement
chemical potential the diquark EoS becomes generally softer than
nuclear matter solution. As in the previous case (in Fig.~\ref{pr0}),
the behavior of the diquark EoS depends on the diquark-diquark
coupling $g$ and consequently on the strength of the diquark
Bose-Einstein condensate. A bigger diquark-diquark coupling $g$ leads
to a stiffer EoS. A diquark-diquark coupling $g$ induces enough
repulsion in order to prevent the collapse of the system to infinite
diquark BEC condensate and makes the system stable. The inserted plot
in Fig.~\ref{pr1} shows more clearly the appearance of several
branches of $P(\mu_{b})$ which is a manifestation of first-order phase
transition. At a fixed baryonic chemical potential only the highest
positive pressure corresponds to a stable phase. In Fig.~\ref{pr1}
(left pannel), we show the baryonic density $\rho_{b}/\rho_{0}$ as a
function of the baryonic chemical potential $\mu_{b}$ for various
parameter sets. The deconfinement phase transition leads to a baryonic
density jump at the critical chemical potential
$\mu_{b}^{\star}=923.25$ MeV similarly to the nuclear matter
liquid-gas phase transition. In this case, the deconfinement
transition can only be first-order since in hadronic side, the
pressure has multi-branches.

\begin{table}
\centering
\begin{tabular}{llll}
\hline
\hline
Parameter ($\mu_{b}^{\star}=923.25$ MeV) &set B2 & set C &set D \\
\hline
$\Lambda_{D}$ (MeV) &650 & 750 & 900 \\
\hline
$g$&16.43 &11.64 & 7.48\\ 
\hline
$g_{d}$&4.56 & 4.85 & 5.1 \\
$m_{d}(\rho=0)$(MeV)&424.1 &451.1& 474.3\\
\hline
\hline
\end{tabular}
\caption{The parameters $\Lambda_{D}$, $g_{d}$ and $g$ in the deconfined phase of the model when 
the critical deconfinement chemical potential $\mu^{\star}_{b}=923.25$
MeV is assumed. Parameter set B2 is already given in table 3. All
parameter sets lead to baryonic density $\rho_{b}^{h}=\rho_{0}$ at the critical deconfinement chemical potential 
$\mu_{b}^{\star}=923.25$ MeV and reproduce the empirical saturation
point ($E_{b}/A=-15.75 ~\text{MeV},\rho_{0}=0.15~\text{fm}^{3}$). }
\end{table}

In order to understand the implication of different choices for the
diquark ultraviolet cutoff $\Lambda_{D}$, we also calculated the
diquark EoS for three different values of the cutoff $\Lambda_{D}$
given in table 4, assuming that the deconfinement takes place at
$\mu^{\star}=923.25$ MeV. All parameter sets given in table 4
reproduced the empirical saturation point
$(E_{b}/A=-15.75~\text{MeV},\rho_{0}=0.15~\text{fm}^{3})$. 
It is seen from table 4 that by increasing the cutoff $\Lambda_{D}$, a
weaker repulsive diquark-diquark coupling $g$ is needed.
\begin{figure}[!tp]
       \centerline{\includegraphics[width=9 cm] {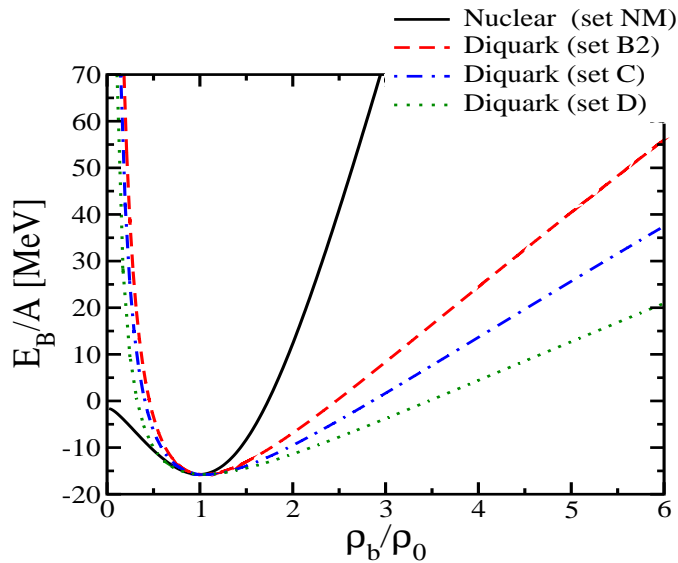}} \caption{The
       binding energy per baryon $E_{B}/A$ as a function of density
       $\rho_{b}/\rho_{0}$ for the nuclear matter with parameter set
       NM (table 1) and for the diquark matter with different diquark ultraviolet cutoff $\Lambda_{D}$ given in table 4.\label{pb}}
\end{figure}
In Fig.~\ref{pb}, we show the binding energy per baryon for the
hadronic matter with parameter set NM (in table 1) and for diquark
matter with various parameter sets given in table 4. For all diquark
parameter sets in table 4, at the critical deconfinement chemical
potential $\mu_{b}^{\star}$ the baryonic density is equal to the
normal nuclear matter density
$\rho_{b}^{d}=\rho_{0}=0.15~\text{fm}^{-3}$.  It is obvious that at
high density diquark matter is energetically favored compared with
nuclear matter while at low density nuclear matter is the stable
one. Note the EoS of the hadron or the diquark EoS alone has a minimum at
the saturation density since we have $P_{h}=P_{q}=0$ at
$\mu_{b}^{\star}=923.25$ MeV, reflecting the thermodynamical
consistency equation (\ref{maxcon}). The produced nuclear-diquark EoS
is softer than the pure nuclear matter EoS\footnote{A computation of
the compression modulus of the joined nuclear-diquark EoS (solid and
dashed lines in Fig.~\ref{pb}) exactly at the saturation point is not
possible, since we have assumed a sharp boundary between the baryonic
and the diquark phases.}. By lowering the diquark ultraviolet cutoff $\Lambda_{D}$, the diquark EoS becomes stiffer, see Fig.~\ref{pb}.

It is plausible that despite the fact that we have turned off the
vector-meson field at the saturation point, the diquark interaction
induces enough repulsion to stabilize matter. In other words, in our
model diquarks play the role of the vector-meson in the deconfined
phase and the rearrangement of quarks from packages of three quarks to
packages of two quarks is essential. In a quark exchange model of
NN-interaction, the s-channel picture of quark exchange looks like two
remaining diquark cores and a diquark in between. Therefore, one could
say that the diquark picture is dual to a meson exchange picture. The
bosonic approach reproduces the correct saturation point and a soft
EoS. In fact, repulsive bosons are easier to handle than fermions
where the correlated wave function cannot be adequately described by
the mean-field theory. Moreover, the diquark BEC goes over directly
into BCS-condensate.

\begin{figure}[!tp]
       \centerline{\includegraphics[width=15 cm] {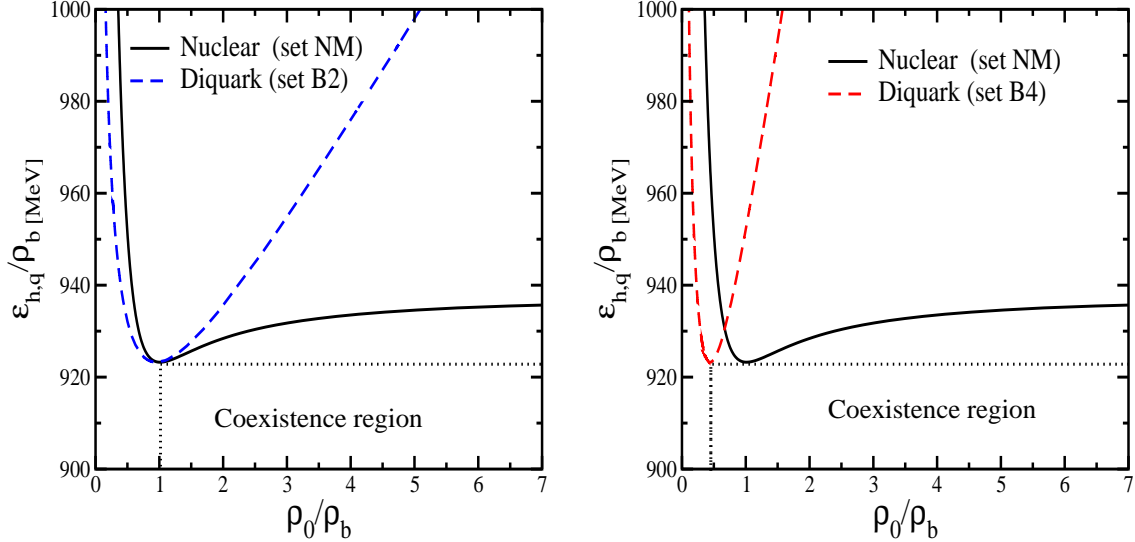}} \caption{
       Energy per baryon number as functions of inverse baryonic
       density $(\rho_{b}/\rho_{0})^{-1}$ in the unit of the nuclear
       matter density $\rho_{0}$. Various parameter sets given in
       tables 1 and 3 are used. The dotted line represents the
       Maxwell-construction which determines the region of
       coexistence: between the nucleons gas, hadronic droplets with
       $\bar{\sigma}\ne 0$, the droplets of diquark Bose-Einstein
       condensate with $\bar{\sigma}=0$ surrounded by a non-trivial
       vacuum with $\bar{\sigma}=93$ MeV. \label{pma1}}
\end{figure}

In Fig.~\ref{pma1}, we show the energy per baryon
$\varepsilon_{h,q}/\rho_{b}$ as a function of inverse baryonic density
$(\rho_{b}/\rho_{0})^{-1}$ for two parameter sets given in tables 1
and 3. They exhibit a typical Van der Waals-like isotherm diagram
where liquid and gaseous phases coexist. At low density in the
confined region, nuclear matter is mechanically unstable since the
pressure is negative. But the pressure in vacuum and at saturation
density $\rho_{0}$ is equal to zero. Therefore, the nuclear matter
mechanical instability can be prevented by fragmenting nuclear matter
into droplets of nuclei, where each droplet has pressure zero and
density equal to $\rho_{0}=0.15~\text{fm}^{-3}$.
Diquark BEC matter also breaks up into stable droplets in which the
pressure is zero and density is $\rho^{d}_{b}$. The diquark BEC
droplets are therefore surrounded by non-perturbative vacuum and
droplets of nucleons.  As we already discussed at the chemical
potential $\mu_{b}=923.25 $ MeV chiral symmetry is restored.
Therefore, diquark droplets exhibit QCD-matter in a chirally restored
form, whereas the chiral symmetry within nucleonic droplets is only
partially restored. The nucleon mass at the saturation density is
about $763$ MeV for the parameter set NM, see Fig. ~\ref{pm1}. Therefore, in our
scenario, multifragmentation has more facets than usually. It also signals a first order phase transition of chiral restoration
and deconfinement.  Ordinary nuclear matter can then be identified as
being in the mixed phase of the first order phase transition:
consisting of droplets of diquarks and droplets of nucleons surrounded
by empty space.

 Within this model we cannot directly identify the diquark BEC
 droplets with abnormal nuclei, since color should be gauged if the
 model is to yield color-singlet droplets. However, one may tempt to
 do so, since we know that within nucleons the chiral symmetry is
 already restored, and the quark density is not zero.  In this case,
 one arrives at a similar picture to that in the MIT bag model. A
 similar picture of a nucleon as a bag within which chiral symmetry is
 restored has also been proposed by Buballa \cite{bub}, Alford,
 Rajagopal and Wilczek \cite{arw} and Berges and Rajagopal \cite{br},
 (see also Ref.~\cite{kl}), but within quite different models and
 mechanisms.  Note that here we can easily choose parameter sets
 where baryon density in droplets of diquarks is different from the
 nucleonic droplet $\rho^{d}_{b}\neq\rho_{0}$. Identifying a diquark
 droplet with abnormal nuclei is possible for $\rho_{b1}>\rho_{0}$,
 this may favor a soft EoS, see Fig.~\ref{pr1}. In this
 picture, it is also possible to have combined droplets of nuclei and
 diquarks. A fraction of baryonic charge can then be delocalized in
 form of a diquark BEC. In this case, diquark concentration within
 nuclei can be smaller than nuclear matter density $\rho_{0}$.
 A possible chemistry of such a mixed matter is beyond the scope of this paper.  
\begin{figure}[!tp]
       \centerline{\includegraphics[width=7 cm] {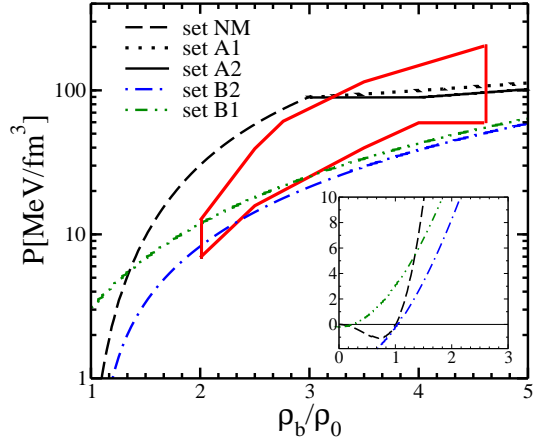}} \caption{
       Pressure as functions of $\rho_{b}/\rho_{0}$ for various
       parameter sets given in tables 1 and 3. Parameter sets A1 and
       A2 (table 2) are given with the assumption that the deconfinement
       takes place at $\mu^{\star}=1214.5$ MeV, corresponding to
       baryonic density $\rho_{b}^{h}/\rho_{0}=3$. Parameter set A2
       leads to a coexistence region between nuclear and chirally
       restored diquark BEC matter within $3\leq\rho_{b}/\rho_{0}\leq
       4$. Parameter sets B1 and B2 (table 3) are given with the assumption
       that the deconfinement takes place right at the nuclear matter
       saturation point $\mu^{\star}_{b}=923.25$ MeV.  The gray closed
       line corresponds to the Danielewicz {\em et al.} constraint
       \cite{ems1}. \label{plast}}
\end{figure}  

There are some experimental restrictions coming from the flow analysis
in heavy-ion collisions which indicates the preferable theoretical
values of pressure in a portion of baryonic density
$2<\rho_{b}/\rho_{0}<4.6 $ at zero temperature \cite{ems1}. This
region is shown in Fig.~\ref{plast} with a gray closed line. It is
important to note that no single EoS can simultaneously describe all
phenomenological data \cite{ems1}. The area of allowed
values does not represent experimental values itself but results from transport calculations for the motions of
nucleons in a collision for a given EoS.  
To compare with the flow constraint
region, we separate regions of SIS ($\rho_{b}<3\rho_{0}$) and AGS
($\rho_{b}>3\rho_{0}$) energies as weak and strong flow
constraints. In this figure, we show the EoS we obtained for two cases
of early deconfinement (sets B1 and B2 in table 3) and late
deconfinement (sets A1 and A2). In both cases, the produced equations
of state pass through the acceptable phenomenological region.  For the
region $\rho_{b}<3\rho_{0}$ the EoS obtained by assuming rather early
deconfinement (sets B1 and B2 in table 2) passes through the
constraint region. These equations of state stay outside the
acceptable phenomenological region at higher density. However, because
of the diquark ultraviolet cutoff, at higher density of about
$\rho_{b}>2.6\rho_{0}$, the model is not trustworthy. The current
available phenomenological data from heavy ion collisions cannot limit
the acceptable boundaries on pressure as a function of density below
two times the normal nuclear matter density
\cite{ems1}.  Note that the models (for sets B1 and B2) assuming the deconfinement phase transition at relatively
low baryonic density cannot be precluded by the present flow
constraint. This is in contrast with conclusion made in
Ref.~\cite{ems1}. For the region $\rho_{b}>3\rho_{0}$ the EoS obtained
by assuming late deconfinement passes through the allowed region.  For
parameter sets A1 and A2, we assumed that the deconfinement takes
place at baryonic density $\rho_{b}=3\rho_{0}$. For parameter set A2,
the deconfinement phase transition is first-order and it leads to a
coexistence region between nuclear and chirally restored diquark BEC
matter within $3<\rho_{b}/\rho_{0}<4$. For parameter set A1, the
deconfinement phase transition is second-order and there is no
coexistence region. Note that since we used rather stiff EoS for the 
hadronic side (set NM), in the confined region the nuclear matter EoS
stays outside allowed region. This is a general problem of these
simple mean-field models and may be improved by adding further meson
interactions or potential terms to the Lagrangian \cite{wal}.

\section{Summary and conclusions}
In this paper, we investigated the nuclear matter and its possible
phase transition to diquark BEC matter at zero temperature. For
the hadronic phase we used a linear $\sigma-\omega$ model incorporating
explicitly the effect of Dirac-sea. For the deconfined phase, we used
a complex scalar field theory in terms of diquarks which couple to
mesonic fields and repel each other. Nucleon, meson and diquark masses
are dynamically generated via spontaneous symmetry breaking.

Our simplified theory does not yet give uniquely the order of a
possible deconfinement transition, nor its location. However, as
pointed out, there are indications which suggest that it might be
first order and takes places above nuclear matter saturation
density. In this paper, we studied the implication of different
choices for the critical deconfinement chemical potential. Our
knowledge about the form of matter above deconfinement is still very
rudimentary. In this paper, we studied a possible formation of a
diquark BEC phase with deconfinement and investigated its implication
for the baryonic matter.  Since we do not know the strength of the
diquark-diquark interaction coupling $g$, we investigated possible
values for this coupling and possible physical consequences.

We found that it is possible that nuclear matter undergoes a diquark BEC
phase transition. Our study generally indicates the importance of the
diquark BEC phenomenon at a rather low chemical potential.  It has been
shown within various models that phase transition from hadronic matter
to normal quark matter cannot occur \cite{bub0,bec-n0}.  Based on our
studies, deconfinement to diquark BEC matter seems to be a viable
option. 

 We showed that chiral phase transition is of first order and
 coincides with deconfinement transition. The compressibility of the
 EoS depends on the diquark-diquark coupling and the critical
 deconfinement chemical potential. We found that an early
 deconfinement and a weaker diquark-diquark interaction (or a stronger
 diquark BEC) soften the EoS. Assuming that deconfinement already
 takes place at saturation, we proposed a scenario in which the
 saturation properties can be described by the diquark BEC
 phenomenon. In contrast to the standard picture, the vector-meson is
 not needed and the diquark-diquark interactions provide enough
 repulsion in order to prevent the collapse of the system. The EoS
 obtained in this way are rather soft.

In particular, we studied the various coexistence patterns which may
emerge if deconfinement takes place at different critical chemical
potential. We showed if deconfinement takes place at higher baryonic
chemical potential, due to a first-order phase transition there exists
a coexistence region between the hadronic and the diquark matter. The
extension of this region in density depends on the diquark
interaction, a weaker diquark-diquark interaction widens this
region. For model parameters with heavy diquark in vacuum (albeit
within baryons), phase transition to diquark matter becomes of
second-order. We also showed that if deconfinement takes place at
about the nuclear matter saturation chemical potential, the ordinary
nuclear matter can then be conceived as being in the mixed phase of
the first order phase transition, consisting of droplets of nucleons
and droplets of chirally restored BEC diquarks surrounded by a 
non-perturbative vacuum.

Diquark Bose-Einstein condensation guarantees a substantial presence
of diquarks in the plasma.  A heavy quark-antiquark pair $(Q\bar{Q})$ may be
converted into a heavy baryon ($Q$-diquark) and exotic unstable
pentaquark ($\bar{Q}$-diquark-diquark) in the presence of the diquark
BEC. Also the ratio of $\Lambda_{c}/\Sigma_{c}$ can be altered in presence
of the diquark BEC, since $\Lambda_{c}$ contains a scalar diquark
while the production of $\Sigma_{c}$ remains unaffected by the
presence of the scalar diquarks \cite{sidi}. It will be challenging for
incoming facility at GSI/FAIR to probe a possible signature for the
diquark BEC matter.  Moreover, the diquark BEC may alter the compact
star cooling behaviour.  In principle, two quarks can be converted
into a diquark and a pair of neutrinos via the following weak
interaction process $u+d \to [ud]_{\text{diquark}}+\bar{\nu}_{e}+\nu_{e}$. These extra neutrinos will be
strongly enhanced in the presence of the diquark BEC. This mechanism
accelerates the cooling of compact star. On the other hand, it has
been shown that the cooling process will be suppressed in the presence
of BCS-pairing \cite{nspr}. Therefore, the cooling behaviour of compact
star may provide a possible signature of existence of the diquark BEC
matter and BEC-BCS crossover.

It is highly desirable to improve our knowledge about the nature of
diquark interaction in medium and its possible connections to QCD
parameters or some phenomenological inputs. It is also of interest to
investigate the implication of the nuclear-diquark model presented
here under neutron stars conditions.

\begin{ack}
A.H.R. would like to thank I. A. Shovkovy for very fruitful discussions
and acknowledges the financial support from the Alexander von Humboldt
foundation.
\end{ack}


\begin{thebibliography}{99}
\bibitem{la1} J. B. Kogut {\em et. al.},  Phys. Rev. Lett. {\bf 50}, 393 (1983); 
F. Karsch and E. Laermann, Phys. Rev. {\bf D50}, 6954 (1994);
F. Karsch, E. Laermann and A. Peikert, Nucl. Phys. {\bf B605}, 579 (2001).
\bibitem{la3}
 O. Philipsen, PoS {\bf LAT2005}, 016 (2005), [hep-lat/0510077].
\bibitem{nqd}
 M. A. Halasz, A. D. Jackson, R. E. Shrock, M. A. Stephanov, J. J. M. Verbaarschot, Phys. Rev. {\bf D58}, 096007 (1998).
\bibitem{de0}
K. Rajagopal and F. Wilczek, hep-ph/0011333; M. G. Alford, Ann. Rev. Nucl. Part. Sci. {\bf 51}, 131 (2001), [hep-ph/0102047].
\bibitem{cpt}
N. Kaiser, S. Fritsch and W. Weise, Nucl. Phys. {\bf A697}, 255 (2002).
\bibitem{res}
 D. H. Rischke, D. T. Son and M. A. Stephanov, Phys. Rev. Lett. {\bf 87}, 062001 (2001).
\bibitem{dm0}
M. Anselmino {\em et. al.}, Rev. Mod. Phys. {\bf 65}, 1199 (1993);
F. Wilczek, hep-ph/0409168. 
\bibitem{n-d}
For example: T. Schafer, E. V. Shuryak and J. J. M. Verbaarschot, Nucl. Phys. {\bf B412} 143 (1994), [hep-ph/9306220];
A. Buck, R. Alkofer, H. Reinhardt, Phys. Lett. {\bf B286}, 29 (1992);
N. Ishii, W. Bentz, K. Yazaki, Nucl. Phys. {\bf A587}, 617 (1995);
A. H. Rezaeian, N. R. Walet and  M. C. Birse, Phys. Rev. {\bf C70}, 065203 (2004), [hep-ph/0408233]; 
A. H. Rezaeian, hep-ph/0507304 (and references therein); 
M. Oettel, G. Hellstern, R. Alkofer and H. Reinhardt, Phys. Rev. {\bf C58}, 2459 (1998), [nucl-th/9805054].
\bibitem{rp}
A. H. Rezaeian and H. J. Pirner, Nucl. Phys. {\bf A769}, 35 (2006), [nucl-th/0510041].
\bibitem{bw}
W. Bentz and A. W. Thomas, Nucl. Phys. {\bf A696}, 138 (2001), [nucl-th/0105022].
\bibitem{arw}
 M. Alford, K. Rajagopal and F. Wilczek, Phys. Lett. {\bf B422}, 247 (1998), [hep-ph/9711395].
\bibitem{br}
J. Berges and K. Rajagopal, Nucl. Phys. {\bf B538}, 215 (1999), [hep-ph/9804233].
\bibitem{bub0}
M. Buballa, Phys. Rept. {\bf 407}, 205 (2005), [hep-ph/0402234].
\bibitem{shov}
D. H. Rischke, Prog. Part. Nucl. Phys. {\bf 52}, 197 (2004), [nucl-th/0305030];
I. A. Shovkovy, Found. Phys. {\bf 35}, 1309 (2005), [nucl-th/0410091].  
\bibitem{bec2}
For example: C. A. Regal, M. Greiner, and D. S. Jin, Phys. Rev. Lett. \textbf{92}, 040403 (2004).
\bibitem{bec1}
M. Matsuzaki, Phys. Rev. {\bf D62}, 017501 (2000), [hep-ph/9910541]; 
E. Babaev, Int. J. Mod. Phys. \textbf{A16}, 1175 (2002), [hep-th/9909052];
M. Kitazawa, T. Koide, T. Kunihiro and Y. Nemoto, Phys. Rev. {\bf D65}, 091504 (2002), [nucl-th/0111022];
H. Abuki, T. Hasuda and K. Itakura, Phys. Rev. \textbf{D65}, 074014 (2002), [hep-ph/0109013];
M. Kitazawa, T. Koide, T. Kunihiro and Y. Nemoto, Phys. Rev. {\bf D70}, 056003 (2004), [hep-ph/0309026];
Y. Nishida and H. Abuki, Phys. Rev. {\bf D72}, 096004 (2005), [hep-ph/0504083];
M. Kitazawa, T. Koide, T. Kunihiro and Y. Nemoto, Prog. Theor. Phys. {\bf 114}, 117 (2005); 
K. Nawa, E. Nakano and H. Yabu, hep-ph/0509029;  H. Abuki, hep-ph/0605081. 
\bibitem{wbec}
J. F. Donoghue and K. S. Sateesh, Phys. Rev. {\bf D38}, 360 (1988);
D. Kastor and J. H. Traschen, Phys. Rev. {\bf D44}, 3791 (1991);
J. E. Horvath, J. A. de Freitas Pacheco and J. C. N. de Araujo, Phys. Rev. {\bf D46}, 4754 (1992).
\bibitem{shv}
V. A. Miransky and I. A. Shovkovy, Phys. Rev. Lett. {\bf 88}, 111601 (2002);
T. Schaefer, D. T. Son, M. A. Stephanov, D. Toublan and J. J. M. Verbaarschot, Phys. Lett. {\bf B522}, 67 (2001);
 D. Blaschke, D. Ebert, K. G. Klimenko, M. K. Volkov and V. L. Yudichev, Phys. Rev. {\bf D70}, 014006 (2004). 
\bibitem{nc}
H. B. Nielsen and S. Chadha, Nucl. Phys. {\bf B105}, 445 (1976).
\bibitem{lc}
 E. M.~Lifshitz and L.P.~Pitaevskii, {\sl Statistical Physics, Part 2}, (Pergamon, 1980).
\bibitem{fs}
R. J. Furnstahl, B. D. Serot and H.-B. Tang, Nucl. Phys. {\bf A618}, 446 (1997).
\bibitem{la2}
A. Manohar and H. Georgi, Nucl. Phys. {\bf B234}, 189 (1984). 
\bibitem{wal}
B. D. Serot and J. D. Walecka, Int. J. Mod. Phys. {\bf E6}, 515 (1997).
\bibitem{me0}
A. H. Rezaeian, nucl-th/0512027.
\bibitem{em1}
V. Koch, T. S. Biro, J. Kunz and U. Mosel, Phys. Lett. {\bf B185}, 1 (1987); 
T. J. Burvenich and D. G. Madland, Nucl. Phys. {\bf A729}, 769 (2003). 
\bibitem{bog}
J. Boguta, Phys. Lett. {\bf B120}, 34 (1983); {\bf 128B}, 19 (1983). 
\bibitem{dm}
M. Hess, F. Karsch, E. Laermann and I. Wetzorke, Phys. Rev. {\bf D58}, 111502 (1998).
\bibitem{ems}
J. P. Blaizot, Phys. Rept. {\bf 64}, 171 (1980);
J. Piekarewicz, Phys. Rev. {\bf C69}, 041301 (2004).
\bibitem{ems1}
 P. Danielewicz, R. Lacey and W. G. Lynch, Science {\bf 298},1592 (2002).
\bibitem{bub}
M. Buballa, Nucl. Phys. {\bf A611}, 393 (1996).
\bibitem{kl}
T. M. Schwarz, S. P. Klevansky and G. Papp, Phys. Rev. {\bf C60}, 055205 (1999).
\bibitem{la4}
P. de Forcrand and S. Kratochvila, Nucl. Phys. (Proc. Suppl.) {\bf B153}, 62 (2006), [hep-lat/0602024];
S. Kratochvila and P. de Forcrand, PoS {\bf LAT2005},  167 (2006),  [hep-lat/0509143]. 
\bibitem{bec-n0}
For example: I. N. Mishustin, L. M. Satarov, H. Stoecker and W. Greiner, Phys. Rev. {\bf C66}, 015202 (2002). 
\bibitem{sidi}
K. S. Sateesh, Phys. Rev. {\bf D45}, 866 (1992).
\bibitem{nspr}
For example: M. G. Alford, J. A. Bowers and K. Rajagopal, J. Phys. {\bf G27}, 541 (2001), [hep-ph/0009357].






\end{thebibliography}
\end{document}